\documentclass[amsmath,amssymb,11pt,superscriptaddress,reprint, preprintnumbers, notitlepage,aps,prl,twocolumn]{revtex4-1}

\pdfoutput=1 

\usepackage[utf8]{inputenc}
\usepackage[english]{babel}
\usepackage{amsmath}
\usepackage{graphicx}
\usepackage{dcolumn}
\usepackage{epsfig}
\usepackage{slashed}
\usepackage{amssymb}
\usepackage{color}
\usepackage[font=small]{caption}
\usepackage[font=small]{subcaption}
\usepackage{url}
\definecolor{MyDarkBlue}{rgb}{0.1, 0.1, 0.8} 
\definecolor{SBlue}{rgb}{0.2, 0.4, 0.7} 
\definecolor{MyLightBlue}{rgb}{0.22,0.51,0.9}
\definecolor{MyGreen}{rgb}{0.0, 0.5, 0.0}
\definecolor{BrickRed}{rgb}{0.8, 0.25, 0.33}
\RequirePackage{hyperref}
\hypersetup{colorlinks, citecolor=BrickRed,linkcolor=MyDarkBlue, urlcolor=MyLightBlue}

\makeatletter

\makeatletter
\renewcommand\@makecaption[2]{%
  \par
  \vskip\abovecaptionskip
  \begingroup
   \small\rmfamily
    \begingroup
     \samepage
     \flushing
     \let\footnote\@footnotemark@gobble
     \@make@capt@title{#1}{#2}\par
    \endgroup
  \endgroup
  \vskip\belowcaptionskip
}
\makeatother
\begin{document}
\preprint{OSU-HEP-19-05}

\title{Zee-Burst:   
A New Probe of Neutrino Non-Standard Interactions at IceCube}

\author{K. S. Babu}
\affiliation{Department of Physics, Oklahoma State University, Stillwater, OK, 74078, USA}

\author{P. S. Bhupal Dev} 
\affiliation{Department of Physics and McDonnell Center for the Space Sciences, Washington University, St. Louis, MO 63130, USA}

\author{Sudip Jana}
\affiliation{Department of Physics, Oklahoma State University, Stillwater, OK, 74078, USA}

\author{Yicong Sui}   
\affiliation{Department of Physics and McDonnell Center for the Space Sciences, Washington University, St. Louis, MO 63130, USA}

\begin{abstract}
    We propose a new way to probe non-standard interactions (NSI) of neutrinos with matter using the ultra-high energy  (UHE) neutrino data at current and future neutrino telescopes. We consider the Zee model of radiative neutrino mass generation as a prototype, which allows two charged scalars -- one $SU(2)_L$-doublet and one singlet, both being leptophilic,  to be as light as 100 GeV, thereby inducing potentially observable NSI with electrons. We show that these light  charged Zee-scalars could give rise to a Glashow-like resonance feature in the UHE neutrino event spectrum at the IceCube neutrino observatory and its high-energy upgrade IceCube-Gen2, which can probe a sizable fraction of the allowed NSI parameter space. 
\end{abstract}

\maketitle

{\textbf{\textit{Introduction.--}}} 
The observation of ultra-high energy (UHE) neutrinos  at the IceCube neutrino observatory~\cite{Aartsen:2013bka, Aartsen:2013jdh, Aartsen:2014gkd,  Aartsen:2015zva, Aartsen:2017mau, Aartsen:2019kpk} has commenced a new era in neutrino astrophysics. Understanding all aspects of these UHE neutrino events, including their sources, energy flux, flavor composition, propagation, and detection, is of paramount importance to both astrophysics and particle physics communities~\cite{Anchordoqui:2013dnh, Ahlers:2018mkf}. A simple, single-component unbroken power-law flux $\Phi(E_\nu)=\Phi_0(E_\nu/100~{\rm TeV})^{-\gamma}$ gives a reasonably good fit to the high-energy starting event (HESE) component of the IceCube data, with the latest best-fit values of $\Phi_0=(6.45^{+1.46}_{-0.46})\times 10^{-18}\ {\rm GeV}^{-1}{\rm cm}^{-2}{\rm s}^{-1}{\rm sr}^{-1}$ and $\gamma=2.89^{+0.20}_{-0.19}$ at $1\sigma$ significance~\cite{Schneider:2019ayi}. 
Any anomalous features in the observed event spectrum could potentially be used as a probe of fundamental physics.  
One such anomalous feature could be in the form of a new resonance. The purpose of this Letter is to show that such a new resonance can arise naturally in the popular Zee model of radiative neutrino masses~\cite{Zee:1980ai, Babu:2019mfe}, which contains two charged scalars. We refer to this Zee-scalar resonance as the `Zee-burst'.

 Within the SM, the only resonance IceCube is sensitive to is the Glashow resonance~\cite{Glashow:1960zz}, where electron anti-neutrinos hitting the target electrons in ice could produce an on-shell $W$-boson: $\bar{\nu}_e e^-\to W^-\to {\rm anything}$. The energy of the incoming neutrino required to make this resonance happen is fixed at $E_\nu=m_W^2/2m_e= 6.3$ PeV. One candidate Glashow event was identified in a partially-contained PeV event (PEPE) search with deposited energy of $5.9\pm 0.18$ PeV~\cite{Aartsen:2017mau, Taboada:2018}, but has not been included in the event spectrum yet~\cite{Aartsen:2019kpk}. The non-observation of Glashow events might be still consistent with the SM expectations within the error bars, given the uncertainty in the source type ($pp$ versus $p\gamma$), as well as ($\nu_e$, $\nu_\mu$, $\nu_\tau$) flavor composition (1:2:0 vs 0:1:0)~\cite{Bhattacharya:2011qu, Barger:2012mz, Biehl:2016psj, Sahu:2016qet, Sui:2018bbh}. On the other hand, the possibility of observing a $Z$-boson resonance ($Z$-burst) at IceCube due to UHE anti-neutrinos interacting with non-relativistic relic neutrinos~\cite{Weiler:1982qy} is bleak,  as the required incoming neutrino energy in this case turns out to be $E_\nu=m_Z^2/2m_\nu \gtrsim 10^{23}$ eV, well beyond the Greisen–Zatsepin–Kuzmin cut-off energy of $\sim 5\times 10^{19}$ eV for the  UHE cosmic rays~\cite{Greisen:1966jv, Zatsepin:1966jv}--the most likely progenitors of the UHE neutrinos (for related discussion, see Ref.~\cite{Fodor:2001qy}). An interesting alternative is the existence of secret neutrino interactions with a light (MeV-scale) $Z'$~\cite{Araki:2014ona, Araki:2015mya, Kamada:2015era, DiFranzo:2015qea} or light neutrinophilic neutral scalar~\cite{Ioka:2014kca, Ng:2014pca, Ibe:2014pja}, in which case the resonance could again fall in the multi-TeV to PeV range which will be accessible at IceCube. Heavy (TeV-scale) resonances induced by neutrino-nucleon interactions mediated by exotic charged  particles, such as leptoquarks~\cite{Barger:2013pla, Dey:2017ede, Becirevic:2018uab, Dorsner:2019vgp}, or squarks in $R$-parity violating supersymmetry~\cite{Carena:1998gd,Dev:2016uxj, Collins:2018jpg, Dev:2019ekc}. 
 have also been discussed. 
 In this Letter, we propose the possibility of {\it light} charged scalar resonances at IceCube, which are intimately related to neutrino mass generation~\cite{Zee:1980ai}, as well as observable non-standard interactions (NSI)~\cite{Wolfenstein:1977ue} (for a recent update, see Ref.~\cite{Dev:2019anc}). 

As a prototypical example, we take the Zee model~\cite{Zee:1980ai} -- one of the most popular radiative neutrino mass models, which contains an $SU(2)_L$-singlet charged scalar $\eta^\pm$ and an $SU(2)_L$-doublet scalar $H_2$, in addition to the SM-like Higgs doublet $H_1$. The original version of the Zee model~\cite{Zee:1980ai} is fully consistent with neutrino oscillation data~\cite{Herrero-Garcia:2017xdu} (for explicit neutrino mass fits, see Ref.~\cite{Babu:2019mfe}), although the Wolfenstein version of the  model~\cite{Wolfenstein:1980sy} which assumes a $Z_2$ symmetry, thus making the diagonal entries of the neutrino mass matrix vanishing, is excluded by oscillation data~\cite{Koide:2001xy, He:2003ih}. Furthermore, it was pointed out in Ref.~\cite{Babu:2019mfe} that both the singlet and doublet charged scalar components can be as light as $\sim 100$  GeV, while satisfying all existing theoretical and experimental constraints in both charged and neutral scalar sectors. More interestingly, such light charged scalars can  lead to sizable diagonal NSI of neutrinos with electrons, with the maximum allowed values of the NSI parameters $(\varepsilon_{ee},\varepsilon_{\mu\mu},\varepsilon_{\tau\tau})=(8\%, 3.8\%, 43\%)$. We show here that the possibility of having a resonance feature with these light charged Zee-scalars (`Zee burst') provides a new probe of NSI at high-energy IceCube, complementary to the low-energy neutrino oscillation and scattering experiments.  

{\textbf{\textit{Light charged scalars in the Zee model.--}}} In the Higgs basis~\cite{Davidson:2005cw}, only the neutral component of $H_1$ gets a vacuum expectation value $\langle H_1^0\rangle =v\simeq 246.2$ GeV, while $H_2$ is parametrized as $H_2=(H_2^+, (H_2^0+iA^0)/\sqrt 2)$. The charged scalars $\{H_2^+,\eta^+\}$ mix in the physical basis to give rise to the physical charged scalar mass eigenstates 
\begin{eqnarray}
        h^+ & \ = \ & \cos\varphi\, \eta^+ + \sin\varphi\, H_2^+ \, , \nonumber \\
        H^+ &\ = \ & -\sin\varphi \,\eta^+ + \cos\varphi \,H_2^+ \, ,
        \label{eq:charged}
    \end{eqnarray}
with the mixing angle $\varphi$  given by 
    \begin{equation}
        \sin{2\varphi} \ = \ \frac{-\sqrt{2} \ v \mu}{m_{H^+}^2-m_{h^+}^2}~,
        \label{mixphi}
    \end{equation}
    where $\mu$ is the dimensionful coefficient of the cubic term $\mu H_1^iH_2^j\epsilon_{ij}\eta^-$ in the scalar potential, with $\{i,j\}$ being the $SU(2)_L$ indices and $\epsilon_{ij}$ being the $SU(2)_L$ antisymmetric tensor. 
    
    The leptonic Yukawa couplings are given by the Lagrangian 
    \begin{align}
        -{\cal L}_Y \ \supset \ &  f_{\alpha\beta}L_\alpha^i L_\beta^j \epsilon_{ij}\eta^+ + \widetilde{Y}_{\alpha \beta} \widetilde{H}_1^i L^{j}_\alpha   \ell_{\beta}^c \epsilon_{ij} \nonumber \\
        & + Y_{\alpha\beta} \widetilde{H}_2^i L^{j}_\alpha  \ell_{\beta}^c \epsilon_{ij}  + {\rm H.c.} \, ,
        \label{eq:LYuk}
    \end{align}
    where $\{\alpha,\beta\}$ are flavor indices,  $\ell^c$ denotes the left-handed antilepton fields, and $\widetilde{H}_a = i \tau_2 H^ \star_a$ ($a=1,2)$ with $\tau_2$ being the second Pauli matrix. The neutrino mass is generated at one-loop level and is given by 
    \begin{equation}
         M_\nu \ = \ \kappa \, (f M_\ell Y + Y^T M_\ell f^T) \, ,
         \label{nuMass}
    \end{equation}
where  $M_\ell=\widetilde Y v/\sqrt 2$ is the charged lepton mass matrix and $\kappa$ is a one-loop factor given by
    \begin{equation}
          \kappa \ = \ \frac{1}{16 \pi^2} \sin{2 \varphi} \log\left(\frac{m_{h^+}^2}{m_{H^+}^2}\right) \, .
          \label{kfactor}
    \end{equation}
According to Eq.~\eqref{nuMass}, the product of the  Yukawa couplings $f$ and $Y$ is constrained by the neutrino oscillation data, which allows for only one of these couplings to be of order one. We will adopt the choice $Y\sim {\cal O}(1)$ and $f\ll 1$, which maximizes the neutrino NSI in the model~\cite{Babu:2019mfe}. 

For the IceCube phenomenology, we are specifically interested in the light charged scalar scenario. This is confronted with several theoretical and experimental constraints, such as charge breaking minima, electroweak precision tests, charged lepton flavor violation (cLFV), collider constraints from LEP and LHC, lepton universality tests and monophoton constraints. It was shown~\cite{Babu:2019mfe} that both $h^+$ and $H^+$ charged scalars can be as light as 100 GeV, while satisfying all these constraints. The main constraints for light charged scalars come from direct searches at LEP, which are applicable as long as $Y_{\alpha e}\neq 0$ for any flavor $\alpha$. More stringent limits from lepton universality tests in $W$ decays~\cite{LEP:2003aa} will apply if $Y_{ee}\neq 0$, restricting the charged scalars masses to above 130 GeV~\cite{Babu:2019mfe}. In what follows, we will consider the scenario where $Y_{\tau e}\neq 0$ and $Y_{\alpha \tau}\neq 0$ for $\alpha=e$ or $\mu$, which satisfies all constraints for $m_{h^+}=100$ GeV, and at the same time, allows for the largest NSI effect. 

{\textbf{\textit{Signature at IceCube.--}}} Expanding the last term in Eq.~\eqref{eq:LYuk}, we get  
\begin{align}
    {\cal L}_Y \ \supset \ Y_{\alpha\beta}(h^-\sin\varphi + H^-\cos\varphi)\nu_\alpha \ell_\beta^c + {\rm H.c.}
    \label{eq:LYuk1}
\end{align}
For $\beta=e$, this will induce neutrino-electron interactions mediated by the charged scalars $h^-$ and $H^-$. For $E_\nu=m_{h^-(H^-)}^2/2m_e$, this will lead to an $h^-(H^-)$-resonance (Zee-burst) at IceCube. There is no interference with the SM Glashow process (even for $\alpha=e$), because the Zee burst involves only right-handed electrons.  Thus, depending on the mass spectrum of $h^-$ and $H^-$, we would expect either one or two additional resonance peaks in the IceCube energy spectrum. We will consider two benchmark scenarios: (i) $m_{h^-}\approx m_{H^-}$, so that the two peaks are indistinguishable, i.e. contribute to the same energy bin, and (ii) $\Delta m_h\equiv m_{H^-} -m_{h^-}=30$ GeV, so that the two peaks are distinguishable (i.e. their dominant contributions fall in different energy bins). 

To estimate the modification to the event spectrum, we compute the number of events in a given energy bin $i$ as
\begin{align}\label{eq:reconst}
N_i \  = \ T \int d\Omega \int_{E_i^{\rm min}}^{E_i^{\rm max}} dE \sum_{\alpha} \Phi_{\nu_\alpha}(E)  A_{\nu_\alpha}(E,\Omega) \, .
\end{align}
Here $T$ is the exposure time for which we use $T_0=2653$ days, corresponding to 7.5 years of live data taking at IceCube~\cite{Aartsen:2019kpk}; $\Omega$ is the solid angle of coverage and we integrate over the whole sky; $E$ is the electromagnetic-equivalent deposited energy which is an approximately linear function of the incoming neutrino energy~\cite{Palladino:2018evm}; the limits of the energy integration $E_i^{\rm min}$ and $E_i^{\rm max}$ give the size of the $i$th deposited energy bin over which the expected number of events is being calculated;  $\Phi_{\nu_\alpha}(E)$ is the differential astrophysical neutrino+anti-neutrino flux for flavor $\alpha$, for which we use a simple, single-component unbroken power-law, isotropic flux $\Phi(E_\nu)=\Phi_0(E_\nu/E_0)^{-\gamma}$ with the IceCube best-fit values of $\Phi_0=6.45\times 10^{-18}\ {\rm GeV}^{-1}{\rm cm}^{-2}{\rm s}^{-1}{\rm sr}^{-1}$ and $\gamma=2.89$~\cite{Schneider:2019ayi}; and $A_{\nu_\alpha}$ is the effective area per energy per solid angle for the neutrino flavor $\nu_\alpha$, which includes the effective neutrino-matter cross section, number density of target nucleons/electrons and acceptance rates for the shower and track events. In presence of new interactions as in Eq.~\eqref{eq:LYuk1}, only the neutrino-electron cross section gets modified, which in turn affects the effective area. For the SM interactions only, we use the publicly available flavor-dependent effective area integrated over solid angle from Ref.~\cite{Aartsen:2017mau} (for 2078 days of IceCube data), along with a $67\%$ increase in the acceptance (for 2653 days of data)~\cite{Aartsen:2019epb}. In presence of non-SM interactions as in Eq.~\eqref{eq:LYuk1}, we rescale the effective area accordingly by taking the ratio of the cross sections, assuming that the acceptance remains the same. 

\begin{figure}[t!]
\includegraphics[width=0.49\textwidth]{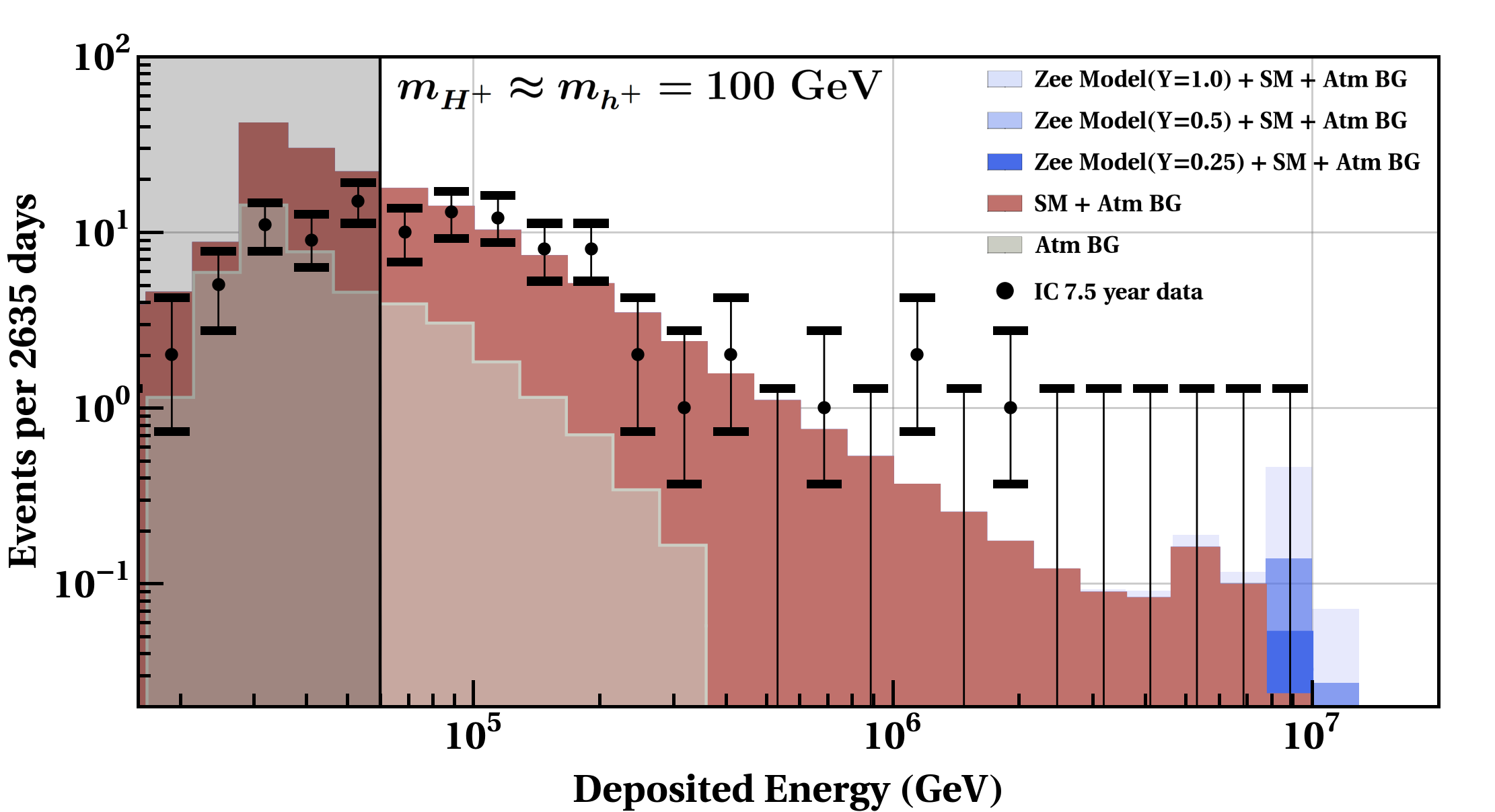}
\caption{Reconstructed event spectra for the expected atmospheric background (gray), SM best-fit with a single-component astrophysical flux (red) and the Zee model with $m_{h^+}\approx m_{H^+}=100$ GeV, $\varphi=\pi/4$ and $Y_{\tau e}=1, 0.5, 0.25$ (light, medium and dark blue, respectively), all compared with the 7.5-year IceCube data. The data points below 60 TeV (inside the vertical black-shaded band) are not included in the IceCube HESE analysis we are using here. \label{fig:spectra}
}
\end{figure}

In the SM, neutrinos interact with nucleons via charged- and neutral-current processes. In the energy range of interest, the corresponding deep inelastic scattering cross sections can be approximated by~\cite{Gandhi:1995tf} 
\begin{align}
\sigma_{\nu(\bar\nu)N}^{\rm CC} \ \approx \  3\sigma_{\nu(\bar\nu) N}^{\rm NC} \ \simeq \  2.7\times 10^{-36} {\rm cm^2}\left(\frac{E_{\nu}}{{\rm GeV}}\right)^{0.4}.
\end{align}
In addition, there are subdominant antineutrino-electron interactions, except in the energy range of 4.6--7.6 PeV, when the $\bar{\nu}_e$--$e^-$ interaction becomes important due to the Glashow resonance~\cite{Glashow:1960zz}.  In the vicinity of the resonance, the dominant piece of the cross section can be expressed by a Breit–Wigner distribution as \cite{Barger:2014iua}:
\begin{align}\label{eq:GlashowXsection}
    \sigma_{\rm Glashow}(s) 
\ = 
\  & 24\pi\,\Gamma_W^2\, {\rm BR}(W^-\to\bar\nu_e e^-){\rm BR}(W^-\to{\rm had}) \nonumber \\
    & \times  \frac{s/m_W^2}{(s-m_W^2)^2+(m_W\Gamma_W)^2} \, ,
\end{align}
where $s=2m_{e}E_{\nu}$ and $\Gamma_{W}$ is the total width of the $W$ boson with ${\rm BR}(W^-\to\bar\nu_e e^-)=10.7\%$ and ${\rm BR}(W^-\to{\rm had})=67.4\%$~\cite{Tanabashi:2018oca}. 
At resonance, Eq.~\eqref{eq:GlashowXsection} gives $\sigma_{\rm Glashow}(E_{\nu}=6.3 \: {\rm PeV})=3.4\times10^{-31}\rm\: cm^2$, about 240 times larger than $\sigma_{\nu(\bar\nu)N}^{\rm CC}(E_{\nu}=6.3\: {\rm PeV})\approx 1.4\times10^{-33}\:\rm cm^2$.
However, due to the narrowness of the resonance and the $E_\nu^{-\gamma}$ nature of the astrophysical neutrino flux, the ratio of the reconstructed events between the resonance-induced $\bar{\nu}_e$-$e$ and non-resonant $\nu(\bar \nu)$-$N$ interactions is not so pronounced in the event spectrum, as shown by the red-shaded histograms in Fig.~\ref{fig:spectra}. 
For instance, for $E_{\nu}>4$ PeV, $N_{\rm Res}/N_{\rm non-Res}\sim 2.05$ giving a total of about 0.3 events in the Glashow bin for the IceCube best-fit flux. Also shown in Fig.~\ref{fig:spectra} (gray shaded) are the total expected atmospheric background (from atmospheric muons and neutrinos, as well as the charm contribution) and the 7.5 year IceCube data~\cite{Schneider:2019ayi}. The vertical line at 60 TeV denotes the low-energy cutoff for the HESE analysis, i.e. the bins below this energy are not considered in the fitting process.  

Now in presence of light charged scalars, we expect a new resonance for $\bar\nu_\alpha e^- \to X^- \to \rm anything$ (where $X^-=h^-,H^-$ for the Zee model) with a cross section similar to Eq~\eqref{eq:GlashowXsection}:
\begin{align}
   \sigma_{\rm Zee} (s) \ = \ & 8\pi\,\Gamma_X^2 \,{\rm BR}(X^-\to\bar\nu_\alpha e^-) {\rm BR}(X^-\to{\rm all}) \nonumber \\
    & \times \frac{s/m_X^2}{(s-m_X^2)^2+(m_X\Gamma_X)^2} \, ,
    \label{eq:ZeeXsection}
\end{align}
where $\Gamma_X=\sum_{\alpha\beta}|Y_{\alpha\beta}|^2{\rm sin^2 \varphi}\:m_X/16\pi$ is the total decay width of $X$. The factor of 1/3, compared to Eq.~\eqref{eq:GlashowXsection}, is due to the difference in the degrees of polarization between scalar and vector bosons. 

In Fig.~\ref{fig:spectra}, we consider a benchmark case with $m_{h^-}\approx m_{H^-}=100$ GeV, so that the two new resonances due to $h^-$ and $H^-$ coincide, and thus, maximize the effect in the bin containing the resonance energy $E_\nu=m_{h^-}^2/2m_e$, as shown by the light, medium and dark blue-shaded histograms corresponding to three illustrative values of $Y_{\tau e}=1,0.5,0.25$ respectively. The excess events due to this new resonance mostly populate the energy bins between 7.6--12.9 PeV, distinguishable from those dominated by the Glashow resonance bin (4.6--7.6 PeV), and the effect is more pronounced for larger Yukawa couplings, as expected from Eq.~\eqref{eq:ZeeXsection}. Here we have taken the maximal mixing $\varphi=\pi/4$ and ${\rm BR}(h^-\to \bar{\nu}_\tau e)=60\%$, ${\rm BR}(h^-\to \bar{\nu}_\beta \tau)=40\%$ (with $\beta=e$ or $\mu$) for a fixed $Y_{\tau e}$ given above and accordingly chosen $Y_{\beta\tau}$, while all other Yukawa couplings $Y_{\alpha\beta}$ are taken to be much smaller than one to satisfy the cLFV  constraints~\cite{Babu:2019mfe}. 
Note that as we increase the mass difference $\Delta m_h\equiv m_{H^-}- m_{h^-}$, the two peaks start populating different bins, but because of the falling power-law flux, the effect is more pronounced in the smallest resonance energy bin. Also note that we cannot make  $\Delta m_h$ exactly zero, otherwise the neutrino mass vanishes [cf.~Eq.~\eqref{kfactor}]. 

\begin{figure*}[t!]
  \centering
 \includegraphics[width=0.49\textwidth]{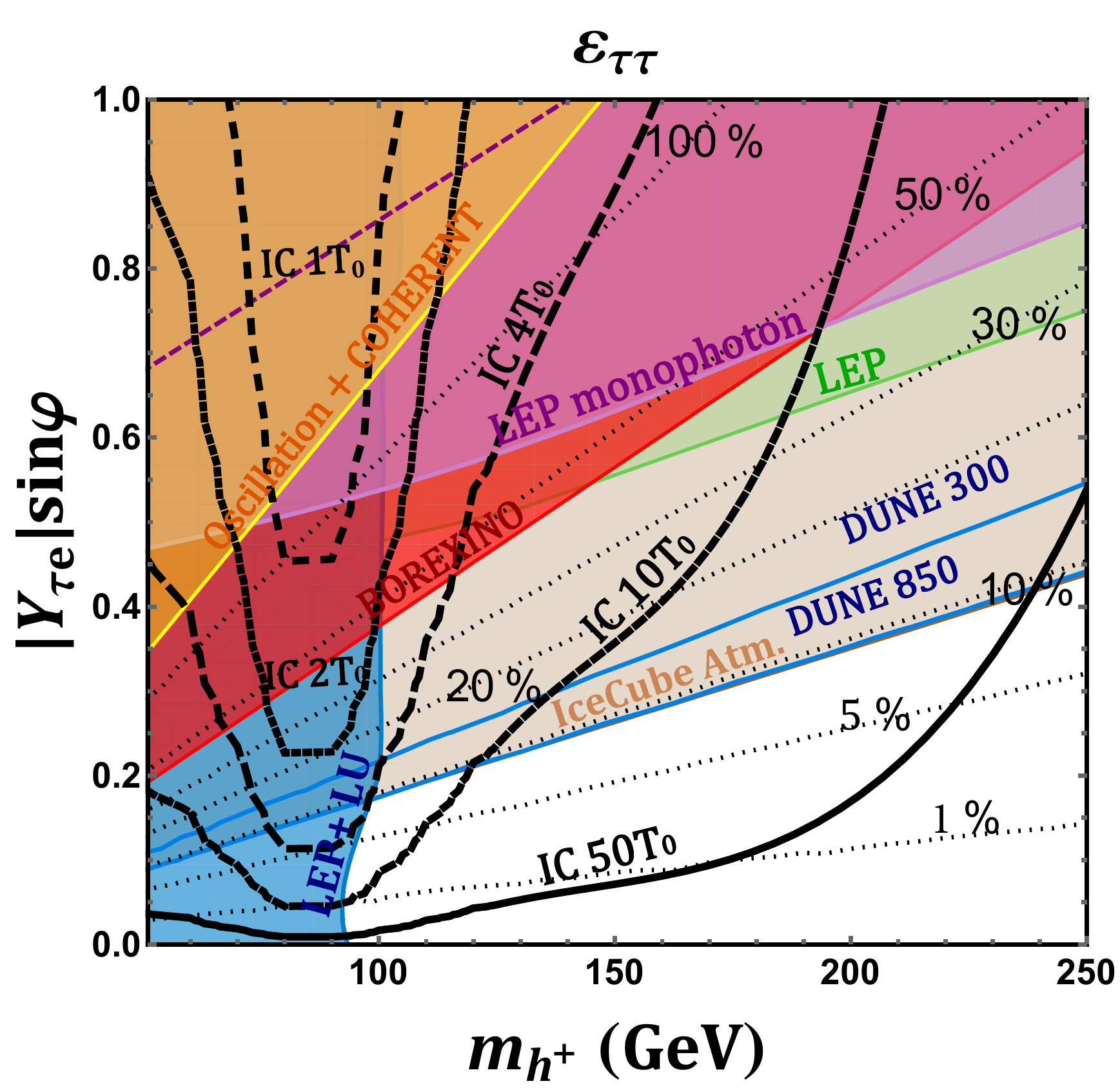}
\includegraphics[width=0.49\textwidth]{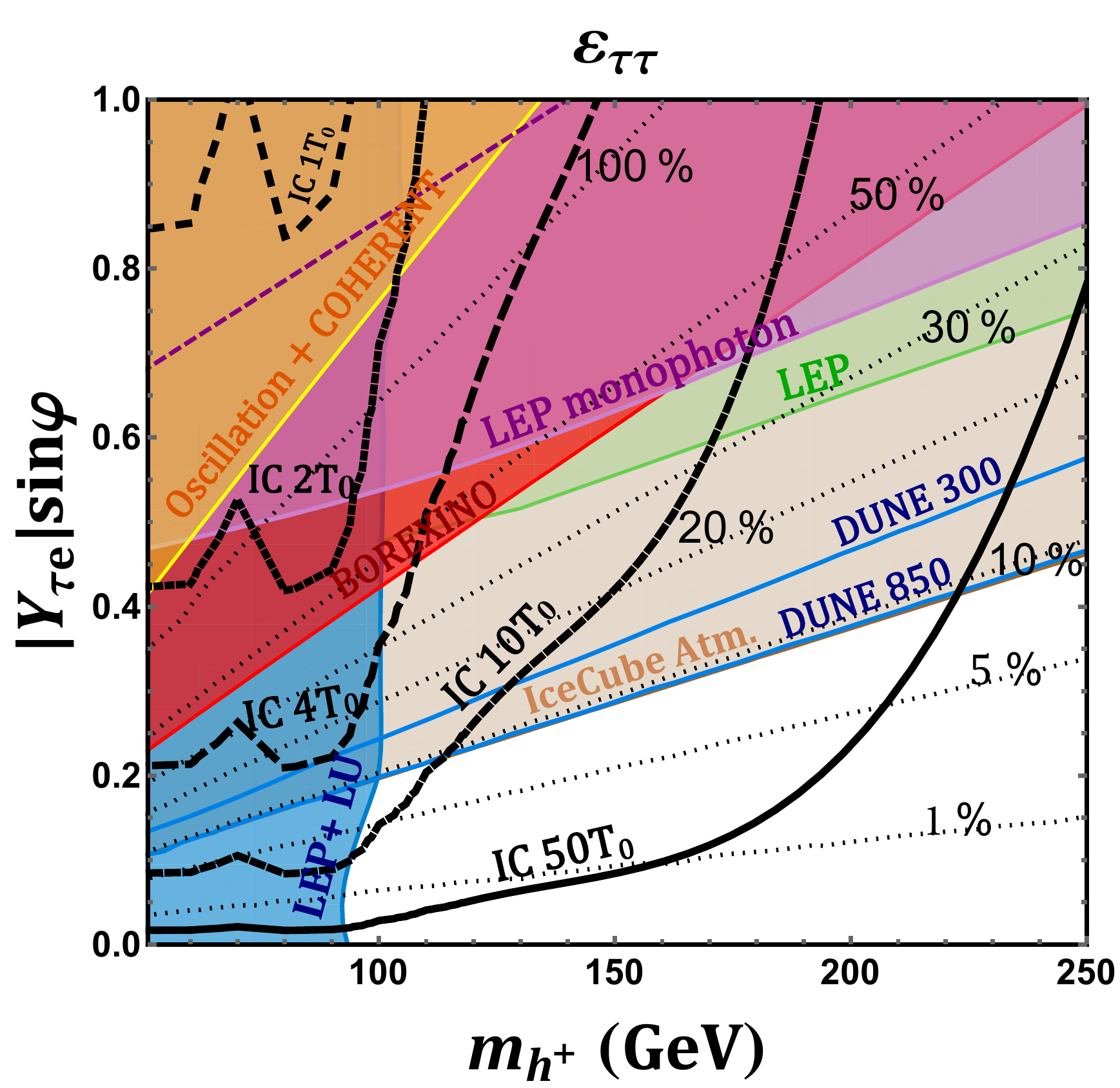}
  \caption{IceCube sensitivity (corresponding to one expected event in the resonance energy bins combined)  for the parameter space relevant for $\varepsilon_{\tau\tau}$ are shown by thick black curves, for different exposure times (in terms of the current exposure $T_0=2653$ days). The left panel is for $m_{h^+}\approx m_{H^+}$ and the right panel is for $m_{H^+}-m_{h^+}=30$ GeV. 
  The predictions for  $\varepsilon_{\tau\tau}$ are shown by the thin dotted contours. The shaded regions are excluded; see text for details. 
  }
  \label{fig:NSI}
\end{figure*}

From Fig.~\ref{fig:spectra}, it is clear that for a given charged scalar mass $m_{h^-}$, the Yukawa coupling $Y_{\tau e}$ cannot be made arbitrarily large without spoiling the best-fit to the observed IceCube HESE data. We can use this fact to derive new IceCube constraints in the $m_{h^-}-Y_{\tau e}$ plane, as shown in Fig.~\ref{fig:NSI} by the thick black contours. The curve labeled `IC 1$T_0$' represents the parameter set which would give rise to one event when summed over the last three bins considered by IceCube best-fit ($4.6<E_\nu/{\rm PeV}<10$) with the current exposure $T_0=2653$ days~\cite{Schneider:2019ayi}, and the other curves are with increased exposures of $2T_0$, $4T_0$, $10T_0$ and $50 T_0$ respectively, keeping the other parameters in Eq.~\eqref{eq:reconst} the same. 
The left panel is for $m_{h+} \approx m_{H^+}$ and the right panel is for $m_{H^+}-m_{h^+}=30$ GeV. This explains the appearance of one `dip' in the left panel (corresponding to one resonance for $h^-$ and $H^-$ combined) and two `dips' in the right panel (corresponding to two distinct resonances for $h^-$ and $H^-$).  

{\textbf{\textit{Probing NSI.--}}} The same Yukawa interactions in Eq.~\eqref{eq:LYuk1} lead to neutrino NSI with electrons, given by~\cite{Babu:2019mfe}
\begin{equation}
    \varepsilon_{\alpha \beta}  \ = \ \frac{Y_{\alpha e } Y_{ \beta e}^{\star}}{4 \sqrt{2}G_F} \left(\frac{ \sin^2{\varphi}}{ m_{h^+}^2}  +   \frac{\cos^2{\varphi} }{m_{H^+}^2}\right) \, ,
         \label{nsieqntot}
\end{equation}
where $G_F$ is the Fermi coupling constant. In Fig.~\ref{fig:NSI}, we show the predictions for $\varepsilon_{\tau\tau}$ by thin black dotted contours. 
Here again we have taken the maximal mixing case with $\varphi=\pi/4$ to get the largest possible NSI. 
The shaded regions are all excluded: blue shaded by direct LEP searches~\cite{lepsusy, Abbiendi:2013hk} and lepton universality (LU) tests in tau decays~\cite{Tanabashi:2018oca}; green shaded by LEP dilepton searches~\cite{LEP:2003aa, Abbiendi:2003dh}; purple shaded  (dashed) by LEP monophoton searches off (on) $Z$-pole~\cite{Acciarri:1998vf, Achard:2003tx}; red shaded by BOREXINO~\cite{Agarwalla:2019smc},  orange shaded by global fit to neutrino oscillation plus COHERENT data~\cite{Esteban:2018ppq}, and brown shaded by IceCube atmospheric neutrino data~\cite{ Esmaili:2013fva, Day:2016shw}. 
For more details on these exclusion regions, see Ref.~\cite{Babu:2019mfe}. Note that the atmospheric neutrino data only constrains $|\varepsilon_{\tau \tau}-\varepsilon_{\mu\mu}|< 9.3\%$~\cite{Esmaili:2013fva, Day:2016shw}, which in the Zee model is equivalent to a bound on $\varepsilon_{\tau\tau}$ itself, because both $\varepsilon_{\tau\tau}$ and  $\varepsilon_{\mu\mu}$ cannot be large simultaneously due to stringent cLFV constraints. One can do similar analysis for other $\varepsilon_{\alpha\beta}$, which are however restricted to be less than a few \%~\cite{Babu:2019mfe}, and hence, are not so promising for IceCube. 

We should comment here that the LEP dilepton constraints~\cite{LEP:2003aa} shown in Fig.~\ref{fig:NSI} (green shaded region) are equally applicable to the extra neutral CP-even and odd scalars ($H,A$) present in the Zee model, since they could modify the $e^+e^-\to \ell_\alpha^+\ell_\alpha^-$ cross section via $t$-channel mediation through the Yukawa couplings $Y_{\alpha e}$. Moreover, these neutral scalars are required to be quasi-degenerate with the doublet charged scalar $H^+$ in order to satisfy the electroweak $T$-parameter constraint~\cite{Babu:2019mfe}.

From Fig.~\ref{fig:NSI}, we see that the existing constraints on NSI are stronger than the current sensitivity of high-energy IceCube data. However, the (non)observation of a resonance-like feature in the future IceCube HESE data could provide a complementary probe of the allowed NSI parameter space, which can even supersede the future DUNE sensitivities (shown by the upper and lower blue solid lines for 300 and 850 kt.MW.yr exposures, respectively~\cite{dev_pondd}). We note here that an exposure of $10T_0$ does not necessarily require 75 years of IceCube running, as a number of factors could improve the conservative projected IceCube limits shown here in a non-linear fashion. For instance, the future data in all the bins may not scale proportionately to the current data and may turn out to be in better agreement with the SM prediction, thus restricting even further any room for new physics contribution. Similarly, the energy-dependent acceptance rate might improve in the future (as it did by 67\% from two to seven years of data~\cite{Aartsen:2019epb}), thereby increasing the effective area, and hence, the `effective' exposure time defined here at a rate faster than linear. Finally, the proposed IceCube-Gen2 with 10 km$^3$ detector  volume~\cite{Aartsen:2014njl} 
could increase the total effective exposure by about an order of magnitude. At the very least, combining IceCube data with the future KM3NeT data~\cite{Adrian-Martinez:2016fdl} could increase the effective exposure by a factor of two.  

Before concluding, we remark that for heavier charged scalars, the resonance energy will be shifted to higher values at which IceCube will become less sensitive, given an isotropic power-law spectrum.  However, if there exists powerful transient sources of UHE neutrinos, then IceCube, as well as current and next-generation radio-Cherenkov neutrino detectors, such as ARA~\cite{Allison:2011wk}, ARIANNA~\cite{Barwick:2014pca}, ANITA~\cite{Gorham:2019guw},  GNO~\cite{Avva:2016ggs}  and RNO~\cite{Aguilar:2019jay}, could be sensitive to electrophilic charged scalars up to a TeV or so (corresponding to the resonance energy of EeV), as might occur e.g. in left-right symmetric model~\cite{Boyarkin:2017rte}. The possibility of a larger flux at higher energies, together with better energy resolution of the IceCube detectors, might help distinguishing the degenerate versus non-degenerate charged-scalar mass spectrum by exploiting the `dip' features in Fig.~\ref{fig:NSI}.

{\textbf{\textit{Conclusion.--}}} We have proposed a new way to probe light charged scalars using a Glashow-like resonance feature in the ultra-high energy neutrino data at IceCube and its future extensions. The same interactions that lead to the new signature at IceCube also give rise to observable non-standard interactions of neutrinos with matter, so that the UHE neutrinos provide a complementary probe of NSI. Taking the popular Zee model of radiative neutrino mass as a prototypical example, we have provided an explicit realization of this idea.

\begin{acknowledgments}
{\textbf {\textit {Acknowledgments.--}}} We thank  Arman Esmaili, Pavel Fileviez Perez, Michele Maltoni and Jordi Salvado for useful discussions. The work of K.S.B. and S.J. was supported in part by US Department of Energy Grant Number DE-SC 0016013. The work of B.D. and Y.S. was supported by the  US  Department  of  Energy  under Grant No.  DE-SC0017987. This work  was also supported by the Neutrino Theory Network Program under Grant No. DE-AC02-07CH11359.  K.S.B., B.D. and S.J.  thank the Fermilab Theory Group for warm hospitality, where part of this work was done. B.D. and Y.S. also  thank the Department of Physics at Oklahoma State University for warm hospitality, where this work was completed.
 \end{acknowledgments}

\bibliographystyle{utphys}
\bibliography{reference}

\providecommand{\href}[2]{#2}\begingroup\raggedright\begin{thebibliography}{10}

\bibitem{Aartsen:2013bka}
{\bfseries IceCube} Collaboration, M.~G. Aartsen {\em et~al.}, ``{First
  observation of PeV-energy neutrinos with IceCube},''
  \href{http://dx.doi.org/10.1103/PhysRevLett.111.021103}{{\em Phys. Rev.
  Lett.} {\bfseries 111} (2013) 021103},
\href{http://arxiv.org/abs/1304.5356}{{\ttfamily arXiv:1304.5356
  [astro-ph.HE]}}.

\bibitem{Aartsen:2013jdh}
{\bfseries IceCube} Collaboration, M.~G. Aartsen {\em et~al.}, ``{Evidence for
  High-Energy Extraterrestrial Neutrinos at the IceCube Detector},''
  \href{http://dx.doi.org/10.1126/science.1242856}{{\em Science} {\bfseries
  342} (2013) 1242856},
\href{http://arxiv.org/abs/1311.5238}{{\ttfamily arXiv:1311.5238
  [astro-ph.HE]}}.

\bibitem{Aartsen:2014gkd}
{\bfseries IceCube} Collaboration, M.~G. Aartsen {\em et~al.}, ``{Observation
  of High-Energy Astrophysical Neutrinos in Three Years of IceCube Data},''
  \href{http://dx.doi.org/10.1103/PhysRevLett.113.101101}{{\em Phys. Rev.
  Lett.} {\bfseries 113} (2014) 101101},
\href{http://arxiv.org/abs/1405.5303}{{\ttfamily arXiv:1405.5303
  [astro-ph.HE]}}.

\bibitem{Aartsen:2015zva}
{\bfseries IceCube} Collaboration, M.~G. Aartsen {\em et~al.}, ``{The IceCube
  Neutrino Observatory - Contributions to ICRC 2015 Part II: Atmospheric and
  Astrophysical Diffuse Neutrino Searches of All Flavors},'' in {\em
  {Proceedings, 34th International Cosmic Ray Conference (ICRC 2015): The
  Hague, The Netherlands, July 30-August 6, 2015}}.
\newblock
\href{http://arxiv.org/abs/1510.05223}{{\ttfamily arXiv:1510.05223
  [astro-ph.HE]}}.
\newblock

\bibitem{Aartsen:2017mau}
{\bfseries IceCube} Collaboration, M.~G. Aartsen {\em et~al.}, ``{The IceCube
  Neutrino Observatory - Contributions to ICRC 2017 Part II: Properties of the
  Atmospheric and Astrophysical Neutrino Flux},''
\href{http://arxiv.org/abs/1710.01191}{{\ttfamily arXiv:1710.01191
  [astro-ph.HE]}}.

\bibitem{Aartsen:2019kpk}
{\bfseries IceCube} Collaboration, M.~G. Aartsen {\em et~al.}, ``{The IceCube
  Neutrino Observatory -- Contributions to the 36th International Cosmic Ray
  Conference (ICRC2019)},''
\newblock 2019.
\newblock
\href{http://arxiv.org/abs/1907.11699}{{\ttfamily arXiv:1907.11699
  [astro-ph.HE]}}.
\newblock

\bibitem{Anchordoqui:2013dnh}
L.~A. Anchordoqui {\em et~al.}, ``{Cosmic Neutrino Pevatrons: A Brand New
  Pathway to Astronomy, Astrophysics, and Particle Physics},''
  \href{http://dx.doi.org/10.1016/j.jheap.2014.01.001}{{\em JHEAp} {\bfseries
  1-2} (2014) 1--30},
\href{http://arxiv.org/abs/1312.6587}{{\ttfamily arXiv:1312.6587
  [astro-ph.HE]}}.

\bibitem{Ahlers:2018mkf}
M.~Ahlers, K.~Helbing, and C.~Pérez de~los Heros, ``{Probing Particle Physics
  with IceCube},'' \href{http://dx.doi.org/10.1140/epjc/s10052-018-6369-9}{{\em
  Eur. Phys. J.} {\bfseries C78} no.~11, (2018) 924},
\href{http://arxiv.org/abs/1806.05696}{{\ttfamily arXiv:1806.05696
  [astro-ph.HE]}}.

\bibitem{Schneider:2019ayi}
A.~Schneider, ``{Characterization of the Astrophysical Diffuse Neutrino Flux
  with IceCube High-Energy Starting Events},'' in {\em {36th International
  Cosmic Ray Conference (ICRC 2019) Madison, Wisconsin, USA, July 24-August 1,
  2019}}.
\newblock 2019.
\newblock
\href{http://arxiv.org/abs/1907.11266}{{\ttfamily arXiv:1907.11266
  [astro-ph.HE]}}.
\newblock

\bibitem{Zee:1980ai}
A.~Zee, ``{A Theory of Lepton Number Violation, Neutrino Majorana Mass, and
  Oscillation},'' \href{http://dx.doi.org/10.1016/0370-2693(80)90349-4,
  10.1016/0370-2693(80)90193-8}{{\em Phys. Lett.} {\bfseries 93B} (1980) 389}.
[Erratum: Phys. Lett.95B,461(1980)].

\bibitem{Babu:2019mfe}
K.~S. Babu, P.~S.~B. Dev, S.~Jana, and A.~Thapa, ``{Non-Standard Interactions
  in Radiative Neutrino Mass Models},''
\href{http://arxiv.org/abs/1907.09498}{{\ttfamily arXiv:1907.09498 [hep-ph]}}.

\bibitem{Glashow:1960zz}
S.~L. Glashow, ``{Resonant Scattering of Antineutrinos},''
\href{http://dx.doi.org/10.1103/PhysRev.118.316}{{\em Phys. Rev.} {\bfseries
  118} (1960) 316--317}.

\bibitem{Taboada:2018}
{\bfseries IceCube} Collaboration, I.~Taboada,
  \href{http://dx.doi.org/10.5281/zenodo.1286919}{``{A View of the Universe
  with the IceCube and ANTARES Neutrino Telescopes},''} in {\em {Proceedings,
  XXVIII International Conference on Neutrino Physics and Astrophysics
  (Neutrino 2018): Heidelberg, Germany, June 4-9, 2018}}.

\bibitem{Bhattacharya:2011qu}
A.~Bhattacharya, R.~Gandhi, W.~Rodejohann, and A.~Watanabe, ``{The Glashow
  resonance at IceCube: signatures, event rates and $pp$ vs. $p\gamma$
  interactions},'' \href{http://dx.doi.org/10.1088/1475-7516/2011/10/017}{{\em
  JCAP} {\bfseries 1110} (2011) 017},
\href{http://arxiv.org/abs/1108.3163}{{\ttfamily arXiv:1108.3163
  [astro-ph.HE]}}.

\bibitem{Barger:2012mz}
V.~Barger, J.~Learned, and S.~Pakvasa, ``{IceCube PeV Cascade Events Initiated
  by Electron-Antineutrinos at Glashow Resonance},''
  \href{http://dx.doi.org/10.1103/PhysRevD.87.037302}{{\em Phys. Rev.}
  {\bfseries D87} no.~3, (2013) 037302},
\href{http://arxiv.org/abs/1207.4571}{{\ttfamily arXiv:1207.4571
  [astro-ph.HE]}}.

\bibitem{Biehl:2016psj}
D.~Biehl, A.~Fedynitch, A.~Palladino, T.~J. Weiler, and W.~Winter,
  ``{Astrophysical Neutrino Production Diagnostics with the Glashow
  Resonance},'' \href{http://dx.doi.org/10.1088/1475-7516/2017/01/033}{{\em
  JCAP} {\bfseries 1701} (2017) 033},
\href{http://arxiv.org/abs/1611.07983}{{\ttfamily arXiv:1611.07983
  [astro-ph.HE]}}.

\bibitem{Sahu:2016qet}
S.~Sahu and B.~Zhang, ``{On the non-detection of Glashow resonance in
  IceCube},'' \href{http://dx.doi.org/10.1016/j.jheap.2018.01.003}{{\em JHEAp}
  {\bfseries 18} (2018) 1--4},
\href{http://arxiv.org/abs/1612.09043}{{\ttfamily arXiv:1612.09043 [hep-ph]}}.

\bibitem{Sui:2018bbh}
Y.~Sui and P.~S.~B. Dev, ``{A Combined Astrophysical and Dark Matter
  Interpretation of the IceCube HESE and Throughgoing Muon Events},''
  \href{http://dx.doi.org/10.1088/1475-7516/2018/07/020}{{\em JCAP} {\bfseries
  1807} no.~07, (2018) 020},
\href{http://arxiv.org/abs/1804.04919}{{\ttfamily arXiv:1804.04919 [hep-ph]}}.

\bibitem{Weiler:1982qy}
T.~J. Weiler, ``{Resonant Absorption of Cosmic Ray Neutrinos by the Relic
  Neutrino Background},''
\href{http://dx.doi.org/10.1103/PhysRevLett.49.234}{{\em Phys. Rev. Lett.}
  {\bfseries 49} (1982) 234}.

\bibitem{Greisen:1966jv}
K.~Greisen, ``{End to the cosmic ray spectrum?},''
\href{http://dx.doi.org/10.1103/PhysRevLett.16.748}{{\em Phys. Rev. Lett.}
  {\bfseries 16} (1966) 748--750}.

\bibitem{Zatsepin:1966jv}
G.~T. Zatsepin and V.~A. Kuzmin, ``{Upper limit of the spectrum of cosmic
  rays},'' {\em JETP Lett.} {\bfseries 4} (1966) 78--80.
[Pisma Zh. Eksp. Teor. Fiz.4,114(1966)].

\bibitem{Fodor:2001qy}
Z.~Fodor, S.~D. Katz, and A.~Ringwald, ``{Determination of absolute neutrino
  masses from Z bursts},''
  \href{http://dx.doi.org/10.1103/PhysRevLett.88.171101}{{\em Phys. Rev. Lett.}
  {\bfseries 88} (2002) 171101},
\href{http://arxiv.org/abs/hep-ph/0105064}{{\ttfamily arXiv:hep-ph/0105064
  [hep-ph]}}.

\bibitem{Araki:2014ona}
T.~Araki, F.~Kaneko, Y.~Konishi, T.~Ota, J.~Sato, and T.~Shimomura, ``{Cosmic
  neutrino spectrum and the muon anomalous magnetic moment in the gauged
  $L_\mu-L_\tau$ model},''
  \href{http://dx.doi.org/10.1103/PhysRevD.91.037301}{{\em Phys. Rev.}
  {\bfseries D91} no.~3, (2015) 037301},
\href{http://arxiv.org/abs/1409.4180}{{\ttfamily arXiv:1409.4180 [hep-ph]}}.

\bibitem{Araki:2015mya}
T.~Araki, F.~Kaneko, T.~Ota, J.~Sato, and T.~Shimomura, ``{MeV scale leptonic
  force for cosmic neutrino spectrum and muon anomalous magnetic moment},''
  \href{http://dx.doi.org/10.1103/PhysRevD.93.013014}{{\em Phys. Rev.}
  {\bfseries D93} no.~1, (2016) 013014},
\href{http://arxiv.org/abs/1508.07471}{{\ttfamily arXiv:1508.07471 [hep-ph]}}.

\bibitem{Kamada:2015era}
A.~Kamada and H.-B. Yu, ``{Coherent Propagation of PeV Neutrinos and the Dip in
  the Neutrino Spectrum at IceCube},''
  \href{http://dx.doi.org/10.1103/PhysRevD.92.113004}{{\em Phys. Rev.}
  {\bfseries D92} no.~11, (2015) 113004},
\href{http://arxiv.org/abs/1504.00711}{{\ttfamily arXiv:1504.00711 [hep-ph]}}.

\bibitem{DiFranzo:2015qea}
A.~DiFranzo and D.~Hooper, ``{Searching for MeV-Scale Gauge Bosons with
  IceCube},'' \href{http://dx.doi.org/10.1103/PhysRevD.92.095007}{{\em Phys.
  Rev.} {\bfseries D92} no.~9, (2015) 095007},
\href{http://arxiv.org/abs/1507.03015}{{\ttfamily arXiv:1507.03015 [hep-ph]}}.

\bibitem{Ioka:2014kca}
K.~Ioka and K.~Murase, ``{IceCube PeV–EeV neutrinos and secret interactions
  of neutrinos},'' \href{http://dx.doi.org/10.1093/ptep/ptu090}{{\em PTEP}
  {\bfseries 2014} no.~6, (2014) 061E01},
\href{http://arxiv.org/abs/1404.2279}{{\ttfamily arXiv:1404.2279
  [astro-ph.HE]}}.

\bibitem{Ng:2014pca}
K.~C.~Y. Ng and J.~F. Beacom, ``{Cosmic neutrino cascades from secret neutrino
  interactions},'' \href{http://dx.doi.org/10.1103/PhysRevD.90.065035,
  10.1103/PhysRevD.90.089904}{{\em Phys. Rev.} {\bfseries D90} no.~6, (2014)
  065035}, \href{http://arxiv.org/abs/1404.2288}{{\ttfamily arXiv:1404.2288
  [astro-ph.HE]}}.
[Erratum: Phys. Rev.D90,no.8,089904(2014)].

\bibitem{Ibe:2014pja}
M.~Ibe and K.~Kaneta, ``{Cosmic neutrino background absorption line in the
  neutrino spectrum at IceCube},''
  \href{http://dx.doi.org/10.1103/PhysRevD.90.053011}{{\em Phys. Rev.}
  {\bfseries D90} no.~5, (2014) 053011},
\href{http://arxiv.org/abs/1407.2848}{{\ttfamily arXiv:1407.2848 [hep-ph]}}.

\bibitem{Barger:2013pla}
V.~Barger and W.-Y. Keung, ``{Superheavy Particle Origin of IceCube PeV
  Neutrino Events},''
  \href{http://dx.doi.org/10.1016/j.physletb.2013.10.021}{{\em Phys. Lett.}
  {\bfseries B727} (2013) 190--193},
\href{http://arxiv.org/abs/1305.6907}{{\ttfamily arXiv:1305.6907 [hep-ph]}}.

\bibitem{Dey:2017ede}
U.~K. Dey, D.~Kar, M.~Mitra, M.~Spannowsky, and A.~C. Vincent, ``{Searching for
  Leptoquarks at IceCube and the LHC},''
  \href{http://dx.doi.org/10.1103/PhysRevD.98.035014}{{\em Phys. Rev.}
  {\bfseries D98} no.~3, (2018) 035014},
\href{http://arxiv.org/abs/1709.02009}{{\ttfamily arXiv:1709.02009 [hep-ph]}}.

\bibitem{Becirevic:2018uab}
D.~Bečirević, B.~Panes, O.~Sumensari, and R.~Zukanovich~Funchal, ``{Seeking
  leptoquarks in IceCube},''
  \href{http://dx.doi.org/10.1007/JHEP06(2018)032}{{\em JHEP} {\bfseries 06}
  (2018) 032},
\href{http://arxiv.org/abs/1803.10112}{{\ttfamily arXiv:1803.10112 [hep-ph]}}.

\bibitem{Dorsner:2019vgp}
I.~Doršner, S.~Fajfer, and M.~Patra, ``{A comparative study of the $S_1$ and
  $U_1$ leptoquark effects at IceCube},''
\href{http://arxiv.org/abs/1906.05660}{{\ttfamily arXiv:1906.05660 [hep-ph]}}.

\bibitem{Carena:1998gd}
M.~Carena, D.~Choudhury, S.~Lola, and C.~Quigg, ``{Manifestations of R-parity
  violation in ultrahigh-energy neutrino interactions},''
  \href{http://dx.doi.org/10.1103/PhysRevD.58.095003}{{\em Phys. Rev.}
  {\bfseries D58} (1998) 095003},
\href{http://arxiv.org/abs/hep-ph/9804380}{{\ttfamily arXiv:hep-ph/9804380
  [hep-ph]}}.

\bibitem{Dev:2016uxj}
P.~S.~B. Dev, D.~K. Ghosh, and W.~Rodejohann, ``{R-parity Violating
  Supersymmetry at IceCube},''
  \href{http://dx.doi.org/10.1016/j.physletb.2016.08.066}{{\em Phys. Lett.}
  {\bfseries B762} (2016) 116--123},
\href{http://arxiv.org/abs/1605.09743}{{\ttfamily arXiv:1605.09743 [hep-ph]}}.

\bibitem{Collins:2018jpg}
J.~H. Collins, P.~S.~B. Dev, and Y.~Sui, ``{R-parity Violating Supersymmetric
  Explanation of the Anomalous Events at ANITA},''
  \href{http://dx.doi.org/10.1103/PhysRevD.99.043009}{{\em Phys. Rev.}
  {\bfseries D99} no.~4, (2019) 043009},
\href{http://arxiv.org/abs/1810.08479}{{\ttfamily arXiv:1810.08479 [hep-ph]}}.

\bibitem{Dev:2019ekc}
P.~S.~B. Dev, \href{http://dx.doi.org/10.1142/9789813275027_0010}{``{Signatures
  of Supersymmetry in Neutrino Telescopes},''} in {\em Probing Particle Physics
  with Neutrino Telescopes, C. P. de los Heros (ed.), World Scientific,
  Singapore (2020)}, pp.~317--352.
\newblock
\href{http://arxiv.org/abs/1906.02147}{{\ttfamily arXiv:1906.02147 [hep-ph]}}.
\newblock

\bibitem{Wolfenstein:1977ue}
L.~Wolfenstein, ``{Neutrino Oscillations in Matter},''
\href{http://dx.doi.org/10.1103/PhysRevD.17.2369}{{\em Phys. Rev.} {\bfseries
  D17} (1978) 2369--2374}.

\bibitem{Dev:2019anc}
P.~S.~B. Dev {\em et~al.}, ``{Neutrino Non-Standard Interactions: A Status
  Report},'' \href{http://dx.doi.org/10.21468/SciPostPhysProc.2.001}{{\em
  SciPost Phys. Proc.} {\bfseries 2} (2019) 001},
\href{http://arxiv.org/abs/1907.00991}{{\ttfamily arXiv:1907.00991 [hep-ph]}}.

\bibitem{Herrero-Garcia:2017xdu}
J.~Herrero-García, T.~Ohlsson, S.~Riad, and J.~Wirén, ``{Full parameter scan
  of the Zee model: exploring Higgs lepton flavor violation},''
  \href{http://dx.doi.org/10.1007/JHEP04(2017)130}{{\em JHEP} {\bfseries 04}
  (2017) 130},
\href{http://arxiv.org/abs/1701.05345}{{\ttfamily arXiv:1701.05345 [hep-ph]}}.

\bibitem{Wolfenstein:1980sy}
L.~Wolfenstein, ``{A Theoretical Pattern for Neutrino Oscillations},''
\href{http://dx.doi.org/10.1016/0550-3213(80)90004-8}{{\em Nucl. Phys.}
  {\bfseries B175} (1980) 93--96}.

\bibitem{Koide:2001xy}
Y.~Koide, ``{Can the Zee model explain the observed neutrino data?},''
  \href{http://dx.doi.org/10.1103/PhysRevD.64.077301}{{\em Phys. Rev.}
  {\bfseries D64} (2001) 077301},
\href{http://arxiv.org/abs/hep-ph/0104226}{{\ttfamily arXiv:hep-ph/0104226
  [hep-ph]}}.

\bibitem{He:2003ih}
X.-G. He, ``{Is the Zee model neutrino mass matrix ruled out?},''
  \href{http://dx.doi.org/10.1140/epjc/s2004-01669-8}{{\em Eur. Phys. J.}
  {\bfseries C34} (2004) 371--376},
\href{http://arxiv.org/abs/hep-ph/0307172}{{\ttfamily arXiv:hep-ph/0307172
  [hep-ph]}}.

\bibitem{Davidson:2005cw}
S.~Davidson and H.~E. Haber, ``{Basis-independent methods for the
  two-Higgs-doublet model},''
  \href{http://dx.doi.org/10.1103/PhysRevD.72.099902,
  10.1103/PhysRevD.72.035004}{{\em Phys. Rev.} {\bfseries D72} (2005) 035004},
  \href{http://arxiv.org/abs/hep-ph/0504050}{{\ttfamily arXiv:hep-ph/0504050
  [hep-ph]}}.
[Erratum: Phys. Rev.D72,099902(2005)].

\bibitem{LEP:2003aa}
{\bfseries LEP, ALEPH, DELPHI, L3, OPAL, LEP Electroweak Working Group, SLD
  Electroweak Group, SLD Heavy Flavor Group} Collaboration, ``{A Combination of
  preliminary electroweak measurements and constraints on the standard
  model},''
\href{http://arxiv.org/abs/hep-ex/0312023}{{\ttfamily arXiv:hep-ex/0312023
  [hep-ex]}}.

\bibitem{Palladino:2018evm}
A.~Palladino and W.~Winter, ``{A multi-component model for observed
  astrophysical neutrinos},'' \href{http://dx.doi.org/10.3204/PUBDB-2018-01376,
  10.1051/0004-6361/201832731}{{\em Astron. Astrophys.} {\bfseries 615} (2018)
  A168},
\href{http://arxiv.org/abs/1801.07277}{{\ttfamily arXiv:1801.07277
  [astro-ph.HE]}}.

\bibitem{Aartsen:2019epb}
M.~G. Aartsen {\em et~al.}, ``{Search for Sources of Astrophysical Neutrinos
  Using Seven Years of IceCube Cascade Events},''
\href{http://arxiv.org/abs/1907.06714}{{\ttfamily arXiv:1907.06714
  [astro-ph.HE]}}.

\bibitem{Gandhi:1995tf}
R.~Gandhi, C.~Quigg, M.~H. Reno, and I.~Sarcevic, ``{Ultrahigh-energy neutrino
  interactions},'' \href{http://dx.doi.org/10.1016/0927-6505(96)00008-4}{{\em
  Astropart. Phys.} {\bfseries 5} (1996) 81--110},
\href{http://arxiv.org/abs/hep-ph/9512364}{{\ttfamily arXiv:hep-ph/9512364
  [hep-ph]}}.

\bibitem{Barger:2014iua}
V.~Barger, L.~Fu, J.~G. Learned, D.~Marfatia, S.~Pakvasa, and T.~J. Weiler,
  ``{Glashow resonance as a window into cosmic neutrino sources},''
  \href{http://dx.doi.org/10.1103/PhysRevD.90.121301}{{\em Phys. Rev.}
  {\bfseries D90} (2014) 121301},
\href{http://arxiv.org/abs/1407.3255}{{\ttfamily arXiv:1407.3255
  [astro-ph.HE]}}.

\bibitem{Tanabashi:2018oca}
{\bfseries Particle Data Group} Collaboration, M.~Tanabashi {\em et~al.},
  ``{Review of Particle Physics},''
\href{http://dx.doi.org/10.1103/PhysRevD.98.030001}{{\em Phys. Rev.} {\bfseries
  D98} no.~3, (2018) 030001}.

\bibitem{lepsusy}
{\bfseries LEP SUSY Working Group} Collaboration.
  \url{http://lepsusy.web.cern.ch/lepsusy/}.

\bibitem{Abbiendi:2013hk}
{\bfseries ALEPH, DELPHI, L3, OPAL, LEP} Collaboration, G.~Abbiendi {\em
  et~al.}, ``{Search for Charged Higgs bosons: Combined Results Using LEP
  Data},'' \href{http://dx.doi.org/10.1140/epjc/s10052-013-2463-1}{{\em Eur.
  Phys. J.} {\bfseries C73} (2013) 2463},
\href{http://arxiv.org/abs/1301.6065}{{\ttfamily arXiv:1301.6065 [hep-ex]}}.

\bibitem{Abbiendi:2003dh}
{\bfseries OPAL} Collaboration, G.~Abbiendi {\em et~al.}, ``{Tests of the
  standard model and constraints on new physics from measurements of fermion
  pair production at 189-GeV to 209-GeV at LEP},''
  \href{http://dx.doi.org/10.1140/epjc/s2004-01595-9}{{\em Eur. Phys. J.}
  {\bfseries C33} (2004) 173--212},
\href{http://arxiv.org/abs/hep-ex/0309053}{{\ttfamily arXiv:hep-ex/0309053
  [hep-ex]}}.

\bibitem{Acciarri:1998vf}
{\bfseries L3} Collaboration, M.~Acciarri {\em et~al.}, ``{Determination of the
  number of light neutrino species from single photon production at LEP},''
\href{http://dx.doi.org/10.1016/S0370-2693(98)00519-X}{{\em Phys. Lett.}
  {\bfseries B431} (1998) 199--208}.

\bibitem{Achard:2003tx}
{\bfseries L3} Collaboration, P.~Achard {\em et~al.}, ``{Single photon and
  multiphoton events with missing energy in $e^{+} e^{-}$ collisions at LEP},''
  \href{http://dx.doi.org/10.1016/j.physletb.2004.01.010}{{\em Phys. Lett.}
  {\bfseries B587} (2004) 16--32},
\href{http://arxiv.org/abs/hep-ex/0402002}{{\ttfamily arXiv:hep-ex/0402002
  [hep-ex]}}.

\bibitem{Agarwalla:2019smc}
{\bfseries Borexino} Collaboration, S.~K. Agarwalla {\em et~al.},
  ``{Constraints on Non-Standard Neutrino Interactions from Borexino
  Phase-II},''
\href{http://arxiv.org/abs/1905.03512}{{\ttfamily arXiv:1905.03512 [hep-ph]}}.

\bibitem{Esteban:2018ppq}
I.~Esteban, M.~C. Gonzalez-Garcia, M.~Maltoni, I.~Martinez-Soler, and
  J.~Salvado, ``{Updated Constraints on Non-Standard Interactions from Global
  Analysis of Oscillation Data},''
  \href{http://dx.doi.org/10.1007/JHEP08(2018)180}{{\em JHEP} {\bfseries 08}
  (2018) 180},
\href{http://arxiv.org/abs/1805.04530}{{\ttfamily arXiv:1805.04530 [hep-ph]}}.

\bibitem{Esmaili:2013fva}
A.~Esmaili and A.~{\relax Yu}. Smirnov, ``{Probing Non-Standard Interaction of
  Neutrinos with IceCube and DeepCore},''
  \href{http://dx.doi.org/10.1007/JHEP06(2013)026}{{\em JHEP} {\bfseries 06}
  (2013) 026},
\href{http://arxiv.org/abs/1304.1042}{{\ttfamily arXiv:1304.1042 [hep-ph]}}.

\bibitem{Day:2016shw}
{\bfseries IceCube} Collaboration, M.~Day, ``{Non-standard neutrino
  interactions in IceCube},''
\href{http://dx.doi.org/10.1088/1742-6596/718/6/062011}{{\em J. Phys. Conf.
  Ser.} {\bfseries 718} no.~6, (2016) 062011}.

\bibitem{dev_pondd}
P.~S.~B. Dev, ``{NSI and Neutrino Mass Models at DUNE},'' {\em Talk given at
  PONDD 2018} .
  \url{https://indico.fnal.gov/event/18430/session/6/contribution/23/material/slides/0.pdf}.

\bibitem{Aartsen:2014njl}
{\bfseries IceCube} Collaboration, M.~G. Aartsen {\em et~al.}, ``{IceCube-Gen2:
  A Vision for the Future of Neutrino Astronomy in Antarctica},''
\href{http://arxiv.org/abs/1412.5106}{{\ttfamily arXiv:1412.5106
  [astro-ph.HE]}}.

\bibitem{Adrian-Martinez:2016fdl}
{\bfseries KM3Net} Collaboration, S.~Adrian-Martinez {\em et~al.}, ``{Letter of
  intent for KM3NeT 2.0},''
  \href{http://dx.doi.org/10.1088/0954-3899/43/8/084001}{{\em J. Phys.}
  {\bfseries G43} no.~8, (2016) 084001},
\href{http://arxiv.org/abs/1601.07459}{{\ttfamily arXiv:1601.07459
  [astro-ph.IM]}}.

\bibitem{Allison:2011wk}
P.~Allison {\em et~al.}, ``{Design and Initial Performance of the Askaryan
  Radio Array Prototype EeV Neutrino Detector at the South Pole},''
  \href{http://dx.doi.org/10.1016/j.astropartphys.2011.11.010}{{\em Astropart.
  Phys.} {\bfseries 35} (2012) 457--477},
\href{http://arxiv.org/abs/1105.2854}{{\ttfamily arXiv:1105.2854
  [astro-ph.IM]}}.

\bibitem{Barwick:2014pca}
{\bfseries ARIANNA} Collaboration, S.~W. Barwick {\em et~al.}, ``{A First
  Search for Cosmogenic Neutrinos with the ARIANNA Hexagonal Radio Array},''
  \href{http://dx.doi.org/10.1016/j.astropartphys.2015.04.002}{{\em Astropart.
  Phys.} {\bfseries 70} (2015) 12--26},
\href{http://arxiv.org/abs/1410.7352}{{\ttfamily arXiv:1410.7352
  [astro-ph.HE]}}.

\bibitem{Gorham:2019guw}
{\bfseries ANITA} Collaboration, P.~W. Gorham {\em et~al.}, ``{Constraints on
  the ultrahigh-energy cosmic neutrino flux from the fourth flight of ANITA},''
  \href{http://dx.doi.org/10.1103/PhysRevD.99.122001}{{\em Phys. Rev.}
  {\bfseries D99} no.~12, (2019) 122001},
\href{http://arxiv.org/abs/1902.04005}{{\ttfamily arXiv:1902.04005
  [astro-ph.HE]}}.

\bibitem{Avva:2016ggs}
J.~Avva {\em et~al.}, ``{Development Toward a Ground-Based Interferometric
  Phased Array for Radio Detection of High Energy Neutrinos},''
  \href{http://dx.doi.org/10.1016/j.nima.2017.07.009}{{\em Nucl. Instrum.
  Meth.} {\bfseries A869} (2017) 46--55},
\href{http://arxiv.org/abs/1605.03525}{{\ttfamily arXiv:1605.03525
  [astro-ph.IM]}}.

\bibitem{Aguilar:2019jay}
J.~A. Aguilar {\em et~al.}, ``{The Next-Generation Radio Neutrino Observatory
  -- Multi-Messenger Neutrino Astrophysics at Extreme Energies},''
\href{http://arxiv.org/abs/1907.12526}{{\ttfamily arXiv:1907.12526
  [astro-ph.HE]}}.

\bibitem{Boyarkin:2017rte}
O.~M. Boyarkin and G.~G. Boyarkina, ``{Can we observe the new physics
  manifestations by the use of ultra-high energy cosmic neutrinos?},''
\href{http://dx.doi.org/10.1016/j.astropartphys.2017.10.001}{{\em Astropart.
  Phys.} {\bfseries 96} (2017) 18--23}.

\end{thebibliography}\endgroup

\end{document}